\documentclass[fp,twocolumn]{jpsj3}
\usepackage{txfonts}

\usepackage{graphicx}  
\usepackage{xcolor}

\usepackage{float}

\usepackage{mathrsfs}
\usepackage{multirow}

\title{Insights into Magnetic Properties of a Mixed Spin-Electron  Model on Decorated Planar Lattices Composed of Half-filled Trigonal Bipyramids}

\author{Lucia G\'alisov\'a$^\ast$}
\inst{Institute of Manufacturing Management,
      Faculty of Manufacturing Technologies with a seat in Pre\v{s}ov, 
      Technical University of Ko\v{s}ice,
      Bayerova 1, 080 01 Pre\v{s}ov, Slovak Republic} 

\abst{A mixed spin-electron model on decorated planar lattices formed by corner-sharing trigonal bipyramids is exactly solved by imposing three mobile electrons per triangular cluster of each bipyramid by means of a generalized decoration-iteration mapping transformation. It is shown that the ground state of the model involves two spontaneously long-range ordered phases, namely, the classical phase with a perfect ferromagnetic arrangement of all particles and the macroscopically degenerate quantum one. The macroscopic degeneracy of the quantum phase lies in chiral degrees of freedom of the mobile electrons. Finite-temperature phase diagrams as well as  temperature variations of the spontaneous magnetization, entropy, and specific heat clearly prove that the classical spin-electron arrangement is thermally almost twice more stable than the quantum one for some combinations of the model's parameters.}

\begin{document}
\maketitle

\section{Introduction}

Mixed spin-electron systems belong to intensively studied materials in today condensed matter physics primarily because the thorough understanding of an origin of their unconventional structural, electronic, magnetic, and other remarkable physical properties opens new ways to many technological applications~\cite{Wach94,Kan63,Tak03,Kam06,Kos12}. It is generally assumed that the aforementioned properties are the result of the mutual interplay between magnetic properties of the systems and the electron motion therein~\cite{Vel88}. However, the relevance of both the contributions is still much discussed.  

Correlated electron systems are usually rigorously unsolvable due to their complexity, and therefore  researchers often resort to some numerical or approximate approaches~\cite{Faz99}. In this regard, exactly solvable mixed spin-electron models are highly desirable, because they represent an excellent playground for a comprehensive  investigation of the impact of electron motion on their magnetic properties at zero and finite temperatures without any artifacts. In the specific case, where the quantum-mechanical hopping of mobile electrons is restricted to finite clusters coupled indirectly through the localized Ising spins, one may adapt the well-known concept of generalized mapping transformations~\cite{Fis59,Syo72,Roj09,Str10} and thus obtain a relevant exact solution of the proposed model. To date, this approach has been successfully applied to various one- (1D)~\cite{Per08,Lis11,Nal14,Gal15a,Gal15b,Str16,Gal17a,Gal17b,Car17,Sou18,Gal18,Car19,Roj21} and two-dimensional (2D)~\cite{Str09,Dor14,Cen16,Cen18a,Cen18b,Cen19,Cen20,Cen21} lattice structures. Despite of their relative simplicity, the investigated mixed spin-electron models have proven to be suitable to simulate many unconventional physical properties and unusual cooperative phenomena with a good qualitative coincidence of the magnetic behavior of real materials. We can mention, for example, the kinetically-driven frustration of the Ising sub-lattice~\cite{Lis11,Nal14,Gal15a,Gal15b,Str16,Gal17a,Gal17b,Gal18,Car19,Roj21}, the local chirality in the electron sub-lattice~\cite{Gal15a,Gal15b,Gal17a,Gal17b,Gal18},  rational~\cite{Per08,Lis11,Nal14,Gal15a,Gal17a,Gal17b,Car19,Roj21,Cen18a} and doping-dependent~\cite{Str16,Gal17a} magnetization plateaus in magnetization curves, double- and also triple-peak temperature dependences of the specific heat~\cite{Per08,Lis11,Nal14,Gal15a,Gal18,Car19}, temperature-induced reentrant phase transitions~\cite{Dor14,Cen16,Cen18a,Cen20}, the bipartite fermionic entanglement between mobile electrons~\cite{Car17,Roj21,Sou18}, and, last but not least, also the enhanced magnetocaloric~\cite{Gal15b,Gal17b} or magnetoelectric~\cite{Cen18b,Cen19,Cen21} effects.

In the present paper we will propose and exactly solve a mixed spin-electron model on decorated planar lattices consisting of corner-sharing trigonal bipyramidal plaquettes. Physically the most interesting half-filled case with three mobile electrons delocalized over the triangular cluster of each plaquette will be particularly investigated in order to bring an insight into differences between phase transitions induced by the classical spontaneous long-range order and quantum spontaneous long-range order with local chiral degrees of freedom of the mobile electrons at zero and also finite temperatures. Besides the academic interest, our theoretical investigation is strongly motivated by existence of the real magnetic copper-based compound Cu$_3$Mo$_2$O$_9$, which can be viewed as the experimental realization of the analogous double-tetrahedral spin chain~\cite{Has08,Kur11,Mat12}. Moreover, there exists a few geometrically frustrated magnetic compounds such as cobaltates $R$BaCo$_4$O$_7$ with a rare-earth atom $R$, in which one can also identify interconnected trigonal bipyramidal units~\cite{Buh14}. Although the latter magnetic compounds do not represent a precise experimental realization of the lattice structure proposed in this paper, their existence indicates the possibility to synthesize real magnetic materials involving interconnected trigonal bipyramidal structures even in higher dimensions. The targeted chemical synthesis involving highly anisotropic spin carriers such Dy$^{3+}$ or Co$^{2+}$ magnetic ions and anion-radical salts could possibly afford the desired mixed spin-electron system. The theoretical findings presented in this paper could serve as a strong motivation for chemists to deal with this issue.

The organization of the paper is as follows. In Sec.~\ref{sec:2} we will introduce the investigated model and the most important steps of its rigorous solution will be explored. The ground-state and finite-temperature phase diagrams along with the typical temperature dependences of the basic thermodynamic quantities will be discussed in detail in Sec.~\ref{sec:3}. Finally, the most interesting findings will be summarized in Sec.~\ref{sec:4}.
\begin{figure}
\vspace{0.5cm}
\centering
\includegraphics[width=1.0\columnwidth]{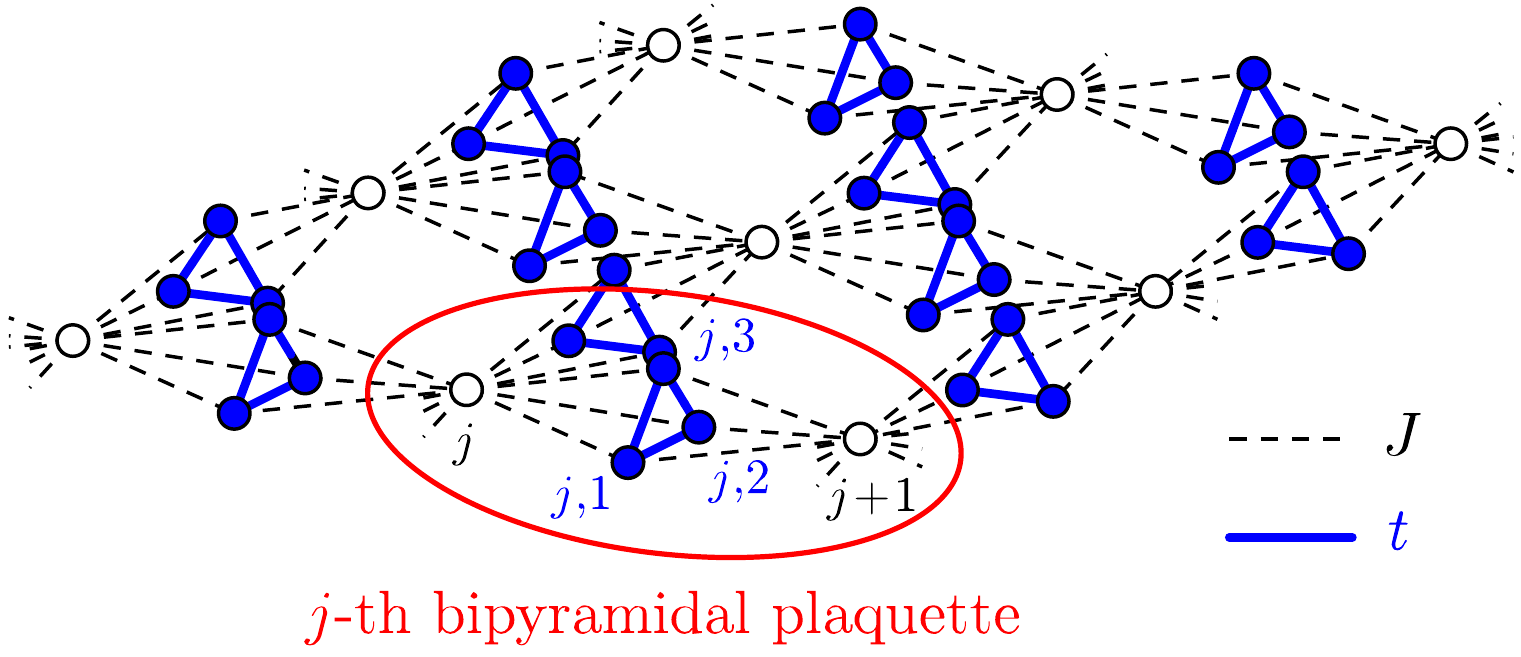}
\caption{(Color online) A schematic representation of the spin-electron model on the bond-decorated square lattice. White circles show nodal lattice sites occupied by the localized Ising spins $\mu = 1/2$, and blue ones forming blue triangular clusters are equivalent interstitial lattice sites available for mobile electrons. 
}
\label{fig:1}
\end{figure}

\section{Model and its exact solution}
\label{sec:2}

We consider a mixed spin-electron model on 2D lattices formed by identical trigonal bipyramidal plaquettes, whose common vertices are occupied by the localized Ising spins of the magnitude $1/2$, while the rest ones, forming equilateral triangular clusters oriented perpendicularly to plaquette axes, are available for mobile electrons. Our attention will be focused on a special half-filled case with exactly three electrons delocalized over the triangular cluster of each plaquette. The quantum-mechanical  hopping of these particles between different trigonal bipyramids is forbidden. For illustration, the particular example of such a lattice with four corner-sharing bipyramidal plaquettes is schematically depicted in Fig.~\ref{fig:1}. Taking into account its special lattice structure, the considered spin-electron system on the lattice with $N$ nodal lattice sites ($N\to\infty$) and $q$ corner-sharing bipyramidal plaquettes may alternatively be viewed as the spin-$1/2$ Ising model on the $q$-coordinated 2D lattice, whose bonds are decorated by half-filled electron triangular clusters. 
From this perspective, the total Hamiltonian of the model can be expressed as a sum of $Nq/2$ plaquette Hamiltonians $\hat{{\cal H}}_j$:
\begin{equation}
\label{eq:H_tot}
\hat{{\cal H}} = \sum_{j=1}^{Nq/2}\hat{{\cal H}}_j,
\end{equation}
where $\hat{{\cal H}}_j$ contains all interaction terms inherent to the $j$-th bipyramidal cell:
{\setlength\arraycolsep{2pt}
\begin{eqnarray}
\label{eq:H_j}
\hat{{\cal H}}_j&=& -t\!\!\sum_{\sigma \in\{\uparrow, \downarrow\}}\sum_{k=1}^3 \big(\hat{c}_{j,k,\sigma}^{\dag}\hat{c}_{j,k+1,\sigma} + \hat{c}_{j,k,\sigma}\hat{c}_{j,k+1,\sigma}^{\dag}\big)
\nonumber\\
&&-\frac{J}{2}\sum_{k = 1}^{3}\,(\hat{n}_{j,k,\uparrow} - \hat{n}_{j,k,\downarrow})(\hat{\mu}_{j}^{z} + \hat{\mu}_{j+1}^{z}) 
+U\sum_{k = 1}^{3}\hat{n}_{j,k,\uparrow}\hat{n}_{j,k,\downarrow}.\quad
\end{eqnarray}}
In above, $\hat{c}_{j,k,\sigma}^{\dag}$ ($\hat{c}_{j,k,\sigma}$) represents fermionic creation (annihilation) operator for mobile electrons with the spin $\sigma \in\{ \uparrow, \downarrow\}$ ($\uparrow$ labels the spin state $1/2$, while $\downarrow$ labels the spin state $-1/2$) at the $k$-th site of the triangular cluster in the $j$-th bipyrapid, $\hat{n}_{j,k,\sigma} = \hat{c}_{j,k,\sigma}^{\dag}\hat{c}_{j,k,\sigma}$ is the fermion number operator, and $\hat{\mu}_{j}^{z}$ is the $z$-component of the spin-$1/2$ operator corresponding to the Ising spin placed at the $j$-th nodal lattice site. The parameter $t>0$ takes into account the kinetic energy of the mobile electrons, $J>0$ ($J<0$) stands for the ferromagnetic (antiferromagnetic) Ising-type interaction between the Ising spins and the nearest-neighboring electrons, and $U>0$ is the on-site Coulomb repulsion between two electrons at the same lattice site. Finally, the periodic boundary conditions $\hat{c}_{j, 4,\sigma}^{\dag} \equiv \hat{c}_{j,1,\sigma}^{\dag}$ ($\hat{c}_{j,4,\sigma} \equiv \hat{c}_{j,1,\sigma}$) and $\hat{\mu}_{N+1}^{z} \equiv \hat{\mu}_{1}^{z}$ within electron triangles and nodal lattice sites are assumed for the sake of simplicity of further calculations.

\subsection{Eigenvalues of the plaquette Hamiltonian}
\label{subsec:2.1}

It is easy to prove that the plaquette Hamiltonian~(\ref{eq:H_j}) remains invariant against various time-independent transformations, namely (a)~the particle-hole transformation $\hat{c}_{j,k,\sigma}^{\dag}\to\hat{c}_{j,k,\sigma}$, $\hat{c}_{j,k,\sigma}\to\hat{c}_{j,k,\sigma}^{\dag}$, (b)~the local cyclic transformation within triangular electron clusters  $\hat{c}_{j,k,\sigma}^{\dag}\to\hat{c}_{j,k+1,\sigma}^{\dag}$ ($\hat{c}_{j,k,\sigma}\to\hat{c}_{j,k+1,\sigma}$), as well as (c)~the translation transformation of nodal lattice sites $\hat{\mu}_{j}^{z}\to \hat{\mu}_{j^{\prime}}^{z}$ ($j\neq j^{\prime}$). Moreover, the total number operator $\hat{n}_{j,\sigma} = \sum_{k=1}^3 \hat{n}_{j,k,\sigma}$, specifying the total number of mobile electrons with the same spin $\sigma$ per half-filled electron triangle, and the total spin operator $\hat{S}_{\!\!j}^z = \sum_{k = 1}^{3}\hat{S}_{\!\!j,k}^z = \sum_{k = 1}^{3}(c_{j,k,\uparrow}^{\dagger}c_{j,k,\uparrow} - c_{j,k,\downarrow}^{\dagger}c_{j,k,\downarrow})/2$, determining the total spin of this cluster in $z$-direction, represent conserved quantities with unambiguously defined quantum numbers $n_{j,\sigma} = \{0,1,2,3\}$ and $S_{\!\!j}^{z} = \{\pm3/2,\pm1/2\}$, respectively. All this implies the validity of the following commutation relations for the plaquette Hamiltonian~(\ref{eq:H_j}): 
\begin{equation}
\label{eq:commutations}
\big[\hat{{\cal H}}_j, \hat{{\cal H}}_{j^{\prime}}\!\big]=0, \quad
\big[\hat{{\cal H}}_j, \hat{n}_{j,\sigma} \big]=0, \quad
\big[\hat{{\cal H}}_j, \hat{S}_{\!\!j}^z \big]=0.
\end{equation}
The first commutation relation listed in Eq.~(\ref{eq:commutations}) clearly indicates that for exact treatment of the considered spin-electron model it is sufficient to find a complete set of eigenvalues of the plaquette Hamiltonian~(\ref{eq:H_j}), because it can be simply extended to the whole lattice. The relevant calculation can be performed in the three-site Hilbert sub-space $\mathscr{H}_{j}$ spanned over the orthonormal basis of twenty possible spin states of the electrons delocalized over the $j$-th triangle of the appropriate plaquette, which can be further divided into four smaller independent subspaces with the fixed eigenvalues $S_{\!\!j}^z$ of the total spin operator $\hat{S}_{\!\!j}^z$ due to validity of the last commutation relation in Eq.~(\ref{eq:commutations}):  
\begin{equation}
\label{eq:Hilbert_j2}
\mathscr{H}_{j} = \mathscr{H}_{S_{\!\!j}^z=-3/2}\oplus\mathscr{H}_{S_{\!\!j}^z=-1/2}\oplus\mathscr{H}_{S_{\!\!j}^z=1/2}\oplus\mathscr{H}_{S_{\!\!j}^z=3/2}. 
\end{equation}
The local cyclic invariance of the three-site electron clusters allows one to write each subspace $\mathscr{H}_{S_{\!\!j}^z}$ as an unification of the electron orbits ${\cal O}_{|f_{j,k}^{ini}\rangle}$ of the cyclic translation group~${\rm C}_3$ with the well-defined initial configuration:
\begin{equation}
\label{eq:fini}
|f_{j,k}^{ini}\rangle = \hat{c}_{j,k,\sigma}^{\dag}\hat{c}_{j,l,\gamma}^{\dag}\hat{c}_{j,p,\tau}^{\dag}|{\rm vac}\rangle\quad (k=1,2,3),  
\end{equation}
where $|{\rm vac}\rangle$ labels the vacuum state. The second and third subscripts in the formula~(\ref{eq:fini}) are given by $l=k+1$, $p=k+2$ and $\sigma = \gamma = \tau = \,\downarrow (\uparrow)$ for $S_{j}^z =-3/2$ ($3/2$), by
$l = (2^k + k - 1)\,{\rm mod}\,3$, $p = (k-3)^2\,{\rm mod}\,3 + l$,
$\sigma = \,\uparrow$ ($\downarrow$) if $k=2$ ($1,3$), $\gamma = \downarrow$ ($\uparrow$) if $l=2$ ($1$), $\tau = S_{j}^z - \sigma - \gamma$ for $ S_{j}^z=-1/2$, and/or by
$\sigma = \,\uparrow$, $\gamma = \,\downarrow$ ($\uparrow$) if $k=2$ ($1,3$), $\tau = S_{j}^z - \sigma - \gamma$ for $ S_{\!\!j}^z=1/2$.
All possible electron configurations constituting individual orbits ${\cal O}_{|f_{j,k}^{ini}\rangle}$ fulfill the following relations:
{\setlength\arraycolsep{2pt}
\begin{eqnarray}
{\rm c}_3 |f_{j,k}^{ini}\rangle &=& \hat{c}_{j,k+1,\sigma}^{\dag}\hat{c}_{j,l+1,\gamma}^{\dag}\hat{c}_{j,p+1,\tau}^{\dag}|{\rm vac}\rangle,
\\
{\rm c}_3^2 |f_{j,k}^{ini}\rangle &=& {\rm c}_3 \!\left({\rm c}_3 |f_{j,k}^{ini}\rangle\right) = 
\hat{c}_{j, k+2,\sigma}^{\dag}\hat{c}_{j,l+2,\gamma}^{\dag}\hat{c}_{j,p+2,\tau}^{\dag}|{\rm vac}\rangle,
\end{eqnarray}}
provided the periodic boundary condition $\hat{c}_{j, 4,\sigma}^{\dag} \equiv \hat{c}_{j,1,\sigma}^{\dag}$. In above, ${\rm c}_3\in{\rm C}_3$ is the cyclic transnational operator. According to the established definitions of ${\cal O}_{|f_{j,k}^{ini}\rangle}$ and $|f_{j,k}^{ini}\rangle$, the sub-spaces $\mathscr{H}_{S_{\!\!j}^z =\mp3/2}$ include just a single basis states with equally oriented spins of all mobile electrons, while $\mathscr{H}_{S_{\!\!j}^z =\mp1/2}$ are spanned over the orthonormal basis of nine different electron states (see Tab.~\ref{tab:1}).
\begin{table*}[h!]
\caption{Decomposition of three electrons over the triangular cluster of the $j$-th bipyramidal plaquette into the orbits ${\cal O}_{|f_{j, k}^{ini}\rangle}$ of the cyclic symmetry group~${\rm C}_{3}$ and the corresponding eigenvalues ${\cal E}_{j,n}$ of the part of the plaquette Hamiltonian~(\ref{eq:H_j}), which described this triangle. The angle $\phi_{U,t}$ in some ${\cal E}_{j,n}$ is the function of two variables $\phi_{x,y} = \frac{1}{3}\arccos\!\Big[x^3/\sqrt{(x^2+27y^2)^3}\Big]$.}
\vspace{2mm}
\label{tab:1}
\centering
\begin{tabular}{rcccl}
\hline\noalign{\smallskip}
$n$ & $S_{\!\!j}^z$ & $|f_{j,k}^{ini}\rangle$ & ${\cal O}_{|f_{j,k}^{ini}\rangle}$ & ${\cal E}_{j,n}$\\
\noalign{\smallskip}
\hline\hline\noalign{\smallskip}
$1$ &
\multirow{1}{*}{$-\frac{3}{2}$} &
        $c_{j,1,\downarrow}^{\dagger}c_{j,2,\downarrow}^{\dagger}c_{j,3,\downarrow}^{\dagger}|{\rm vac}\rangle$ &
        $c_{j,1,\downarrow}^{\dagger}c_{j,2,\downarrow}^{\dagger}c_{j,3,\downarrow}^{\dagger}|{\rm vac}\rangle = |\downarrow,\downarrow,\downarrow\rangle_{j}$ &
        $0$
        \\
  \noalign{\smallskip}\cline{2-5}\noalign{\smallskip}        
$2$ & \multirow{1}{*}{$-\frac{1}{2}$} &
        $c_{j,1,\downarrow}^{\dagger}c_{j,2,\downarrow}^{\dagger}c_{j,3,\uparrow}^{\dagger}|{\rm vac}\rangle$ &
        $c_{j,1,\downarrow}^{\dagger}c_{j,2,\downarrow}^{\dagger}c_{j,3,\uparrow}^{\dagger}|{\rm vac}\rangle= |\downarrow,\downarrow,\uparrow\rangle_{j}$ &
        $0$
        \\
$3$ &   & &
        $c_{j,2,\downarrow}^{\dagger}c_{j,3,\downarrow}^{\dagger}c_{j,1,\uparrow}^{\dagger}|{\rm vac}\rangle= |\uparrow,\downarrow,\downarrow\rangle_{j}$ &
        $U$
        \\
$4$ &   & &
        $c_{j,3,\downarrow}^{\dagger}c_{j,1,\downarrow}^{\dagger}c_{j,2,\uparrow}^{\dagger}|{\rm vac}\rangle= |\downarrow,\uparrow,\downarrow\rangle_{j}$ &
        $U$
        \\	
    \noalign{\smallskip}\cline{3-5}\noalign{\smallskip}
$5$ &   & $c_{j,2,\uparrow}^{\dagger}c_{j,2,\downarrow}^{\dagger}c_{j,3,\downarrow}^{\dagger}|{\rm vac}\rangle$ &
     $c_{j,2,\uparrow}^{\dagger}c_{j,2,\downarrow}^{\dagger}c_{j,3,\downarrow}^{\dagger}|{\rm vac}\rangle= |\emptyset,\uparrow\downarrow,\downarrow\rangle_{j}$ &
     $\frac{2U}{3}-\frac{2}{3}\sqrt{U^2+27t^2}\cos\phi_{U,t}$
        \\
$6$ &   & &
        $c_{j,3,\uparrow}^{\dagger}c_{j,3,\downarrow}^{\dagger}c_{j,1,\downarrow}^{\dagger}|{\rm vac}\rangle= |\downarrow,\emptyset,\uparrow\downarrow\rangle_{j}$ &
     $\frac{2U}{3}+\frac{1}{3}\sqrt{U^2+27t^2}\left(\cos\phi_{U,t} +\sqrt{3}\sin\phi_{U,t}\right)$
        \\
$7$ &   & &
        $c_{j,1,\uparrow}^{\dagger}c_{j,1,\downarrow}^{\dagger}c_{j,2,\downarrow}^{\dagger}|{\rm vac}\rangle= |\uparrow\downarrow,\downarrow,\emptyset\rangle_{j}$ &
     $\frac{2U}{3}+\frac{1}{3}\sqrt{U^2+27t^2}\left(\cos\phi_{U,t} -\sqrt{3}\sin\phi_{U,t}\right)$
        \\
    \noalign{\smallskip}\cline{3-5}\noalign{\smallskip}
$8$ & 	&   $c_{j,3,\downarrow}^{\dagger}c_{j,1,\uparrow}^{\dagger}c_{j,1,\downarrow}^{\dagger}|{\rm vac}\rangle$ &
        $c_{j,3,\downarrow}^{\dagger}c_{j,1,\uparrow}^{\dagger}c_{j,1,\downarrow}^{\dagger}|{\rm vac}\rangle= |\uparrow\downarrow,\emptyset,\downarrow\rangle_{j}$ &
     $\frac{2U}{3}-\frac{2}{3}\sqrt{U^2+27t^2}\cos\phi$
        \\
$9$ &   & &
        $c_{j,1,\downarrow}^{\dagger}c_{j,2,\uparrow}^{\dagger}c_{j,2,\downarrow}^{\dagger}|{\rm vac}\rangle= |\downarrow,\uparrow\downarrow,\emptyset\rangle_{j}$ &
     $\frac{2U}{3}+\frac{1}{3}\sqrt{U^2+27t^2}\left(\cos\phi_{U,t} +\sqrt{3}\sin\phi_{U,t}\right)$
        \\
$10$ &   & &
        $c_{j,2,\downarrow}^{\dagger}c_{j,3,\uparrow}^{\dagger}c_{j,3,\downarrow}^{\dagger}|{\rm vac}\rangle= |\emptyset,\downarrow,\uparrow\downarrow\rangle_{j}$ &
     $\frac{2U}{3}+\frac{1}{3}\sqrt{U^2+27t^2}\left(\cos\phi_{U,t} -\sqrt{3}\sin\phi_{U,t}\right)$
        \\
   \noalign{\smallskip}\cline{2-5}\noalign{\smallskip}       
$11$ &  \multirow{1}{*}{$\frac{1}{2}$} &
        $c_{j,1,\uparrow}^{\dagger}c_{j,2,\uparrow}^{\dagger}c_{j,3,\downarrow}^{\dagger}|{\rm vac}\rangle$ &
        $c_{j,1,\uparrow}^{\dagger}c_{j,2,\uparrow}^{\dagger}c_{j,3,\downarrow}^{\dagger}|{\rm vac}\rangle= |\uparrow,\uparrow,\downarrow\rangle_{j}$ &
        $0$
        \\
$12$ &     & &
        $c_{j,2,\uparrow}^{\dagger}c_{j,3,\uparrow}^{\dagger}c_{j,1,\downarrow}^{\dagger}|{\rm vac}\rangle= |\downarrow,\uparrow,\uparrow\rangle_{j}$ &
        $U$
        \\
$13$ &     & &
        $c_{j,3,\uparrow}^{\dagger}c_{j,1,\uparrow}^{\dagger}c_{j,2,\downarrow}^{\dagger}|{\rm vac}\rangle= |\uparrow,\downarrow,\uparrow\rangle_{j}$ &
        $U$
        \\
	\noalign{\smallskip}\cline{3-5}\noalign{\smallskip}
$14$ & 	 &  $c_{j,2,\uparrow}^{\dagger}c_{j,2,\downarrow}^{\dagger}c_{j,3,\uparrow}^{\dagger}|{\rm vac}\rangle$ &
        $c_{j,2,\uparrow}^{\dagger}c_{j,2,\downarrow}^{\dagger}c_{j,3,\uparrow}^{\dagger}|{\rm vac}\rangle= |\emptyset,\uparrow\downarrow,\uparrow\rangle_{j}$ &
     $\frac{2U}{3}-\frac{2}{3}\sqrt{U^2+27t^2}\cos\phi_{U,t}$
        \\
$15$ &     & &
        $c_{j,3,\uparrow}^{\dagger}c_{j,3,\downarrow}^{\dagger}c_{j,1,\uparrow}^{\dagger}|{\rm vac}\rangle= |\uparrow,\emptyset,\uparrow\downarrow\rangle_{j}$ &
     $\frac{2U}{3}+\frac{1}{3}\sqrt{U^2+27t^2}\left(\cos\phi_{U,t} +\sqrt{3}\sin\phi_{U,t}\right)$
        \\
$16$ &     & &
        $c_{j,1,\uparrow}^{\dagger}c_{j,1,\downarrow}^{\dagger}c_{j,2,\uparrow}^{\dagger}|{\rm vac}\rangle= |\uparrow\downarrow,\uparrow,\emptyset\rangle_{j}$ &
     $\frac{2U}{3}+\frac{1}{3}\sqrt{U^2+27t^2}\left(\cos\phi_{U,t} -\sqrt{3}\sin\phi_{U,t}\right)$
        \\
	\noalign{\smallskip}\cline{3-5}\noalign{\smallskip}
$17$ &     &   $c_{j,3,\uparrow}^{\dagger}c_{j,1,\uparrow}^{\dagger}c_{j,1,\downarrow}^{\dagger}|{\rm vac}\rangle$ &
        $c_{j,3,\uparrow}^{\dagger}c_{j,1,\uparrow}^{\dagger}c_{j,1,\downarrow}^{\dagger}|{\rm vac}\rangle= |\uparrow\downarrow,\emptyset,\uparrow\rangle_{j}$ &
     $\frac{2U}{3}-\frac{2}{3}\sqrt{U^2+27t^2}\cos\phi_{U,t}$
        \\
$18$ &     & &
        $c_{j,1,\uparrow}^{\dagger}c_{j,2,\uparrow}^{\dagger}c_{j,2,\downarrow}^{\dagger}|{\rm vac}\rangle= |\uparrow,\uparrow\downarrow,\emptyset\rangle_{j}$ &
     $\frac{2U}{3}+\frac{1}{3}\sqrt{U^2+27t^2}\left(\cos\phi_{U,t} +\sqrt{3}\sin\phi_{U,t}\right)$
        \\
$19$ &     & &
        $c_{j,2,\uparrow}^{\dagger}c_{j,3,\uparrow}^{\dagger}c_{j,3,\downarrow}^{\dagger}|{\rm vac}\rangle= |\emptyset,\uparrow,\uparrow\downarrow\rangle_{j}$ &
     $\frac{2U}{3}+\frac{1}{3}\sqrt{U^2+27t^2}\left(\cos\phi_{U,t} -\sqrt{3}\sin\phi_{U,t}\right)$
        \\
   \noalign{\smallskip}\cline{2-5}\noalign{\smallskip}   
$20$ &  \multirow{1}{*}{$\frac{3}{2}$} &
        $c_{j,1,\uparrow}^{\dagger}c_{j,2,\uparrow}^{\dagger}c_{j,3,\uparrow}^{\dagger}|{\rm vac}\rangle$ &
        $c_{j,1,\uparrow}^{\dagger}c_{j,2,\uparrow}^{\dagger}c_{j,3,\uparrow}^{\dagger}|{\rm vac}\rangle= |\uparrow,\uparrow,\uparrow\rangle_{j}$ &
        $0$
        \\
\noalign{\smallskip}\hline
\end{tabular}
\end{table*}

The decomposition of the Hilbert electron subspace~(\ref{eq:Hilbert_j2}) implies that the plaquette Hamiltonian~(\ref{eq:H_j}) consists of two one-element and two $9\times9$ disjoint hermitian blocks. The latter ones can be further divided into three smaller disjoint blocks by applying the so-called basis of wavelets~\cite{Lul03,Jak09,Jak13} on the orbits ${\cal O}_{|f_{j,k}^{ini}\rangle}\subset \mathscr{H}_{S_{\!\!j}^z=\mp1/2}$. In the present notation, the appropriate amplitude takes the form:
\begin{equation}
|b_j, f_{j,k}^{ini}\rangle = \frac{1}{\sqrt{3}}\sum_{k=1}^{3}{\rm e}^{2\pi{\rm i}b_jk/3} |f_{j,k}, f_{j,k}^{ini}\rangle,
\end{equation}
where ${\rm i}$ is the imaginary unit satisfying ${\rm i}^2 = -1$, $|f_{j,k}, f_{j,k}^{ini}\rangle$  denotes the $k$-th electron configuration of ${\cal O}_{|f_{j,k}^{ini}\rangle}$, and $b_j$ is the discrete quasi-momentum of the Brillouin zone $B = \{b_j = 0,\pm1\}$. At this stage, the task of searching for a complete set of eigenvalues of the plaquette Hamiltonian~(\ref{eq:H_j}) is divided into eight independent tasks corresponding to individual electron orbits ${\cal O}_{|f_{j,k}^{ini}\rangle}$. Their solutions lead to the following unified analytical expression for the energy spectrum of the Hamiltonian~(\ref{eq:H_j}):
\begin{equation}
\label{eq:E_j}
E_{j,n} = -JS_{\!\!j}^z(\mu_{j}^{z} + \mu_{j+1}^{z}) + {\cal E}_{j,n} \quad(n=1,2,\ldots,20),
\end{equation}
where ${\cal E}_{j,n}$ denotes the appropriate eigenvalue of the half-filled electron triangle in the $j$-th bipyramidal plaquette. Explicit expressions of ${\cal E}_{j,n}$ are listed in Tab.~\ref{tab:1}. The energy spectrum~(\ref{eq:E_j}) can be further employed for a comprehensive ground-state analysis, as well as a rigorous solution of the partition function of the investigated model needed for successful study of its finite-temperature properties. 

\subsection{Partition function}
\label{subsec:2.2}

Because of different plaquette Hamiltonians commute with each other, the partition function ${\cal Z} = \Tr{\rm e}^{-\beta\hat{{\cal H}}} $ of the 2D mixed spin-electron model can be partially factorized and expressed in terms of the eigenenergies of the Hamiltonian~(\ref{eq:H_j}): 
\begin{equation}
\label{eq:Z}
{\cal Z} = \sum_{\{\mu_l\}}\prod_{j=1}^{Nq/2}\Tr_{j}{\rm e}^{-\beta\hat{{\cal H}}_j} = \sum_{\{\mu_l\}}\prod_{j=1}^{Nq/2}\sum_{n=1}^{20}{\rm e}^{-\beta E_{j,n}}.
\end{equation}
In above, $\beta = 1/(k_{\rm B}T)$ ($k_{\rm B}$ is the Boltzmann's constant and $T$ is the absolute temperature of the system), the symbol $\sum_{\{\mu_l\}}$ denotes the summation over all possible spin states of the nodal Ising spins, the product symbol $\prod_{j=1}^{Nq/2}$ runs over bipyramidal plaquettes, and $\Tr_{j}$ stands for the trace over all possible degrees of freedom of three mobile electrons from the $j$-th triangular cluster of the  plaquette. Finally, the summation symbol $\sum_{n=1}^{20}$ behind the second equal symbol counts all eigenvalues of the plaquette Hamiltonian~(\ref{eq:H_j}) given by Eq.~(\ref{eq:E_j}). After performing $\sum_{n=1}^{20}$, one obtains the effective Boltzmann’s weight $w(\mu_j^z,\mu_{j+1}^z)$, whose explicit form gives the opportunity to use the generalized decoration-iteration mapping transformation~\cite{Fis59,Syo72,Roj09,Str10}:
{\setlength\arraycolsep{2pt}
\begin{eqnarray}
\label{eq:DIT}
w(\mu_j^z, \mu_{j+1}^z) &=& \sum_{n=1}^{20}{\rm e}^{-\beta E_{j,n}}  = 2\cosh\left[\frac{3\beta J}{2}(\mu_j^z + \mu_{j+1}^z)\right]\nonumber \\
&&
+2\cosh\left[\frac{\beta J}{2}(\mu_j^z + \mu_{j+1}^z)\right]\!\Bigg\{ 1+2{\rm e}^{-\beta U}
\nonumber \\
&&
+2{\rm e}^{-2\beta\left(U - \sqrt{U^2+27t^2}\,\right)/3}\sum_{r=1}^{3}\cos\left(\frac{\varphi_r}{3}\right)\!\Bigg\}
\nonumber \\
&=& A{\rm e}^{\beta J_{eff}\mu_j^z\mu_{j+1}^z},
\end{eqnarray}}
where $\varphi_r = 2\pi r + \arctan\left[9t\sqrt{(U^2+27t^2)^3}/U^3\right]$.
An essence of the used transformation is to substitute all lattice sites available for the mobile electrons and associated interactions by a novel effective Ising-type coupling $J_{eff}$ between remaining nodal Ising spins. The novel mapping parameters $A$ and $J_{eff}$ emerging in the last line of Eq.~(\ref{eq:DIT}) are determined by 'self-consistency' of the used algebraic technique:
\begin{equation}
\label{eq:AJef}
A =
\sqrt{w_{0}\,w_{1}}\,,\quad
J_{eff} = 2k_{\rm B}T\ln\left(\frac{w_{1}}{w_{0}}\right),
\end{equation}
where the Boltzmann’s weights $w_{0} = w(\pm1/2, \mp1/2)$ and $w_{1} = w(\pm1/2, \pm1/2)$ evidently satisfy the inequality $w_1\geq w_0$. Hence, the effective temperature-dependent coupling $J_{eff}$ remains non-negative over the entire parameter range and after substituting Eq.~(\ref{eq:DIT}) into Eq.~(\ref{eq:Z}) one obtains the following universal rigorous relation between the partition function~${\cal Z}$ of the mixed spin-electron model given by the Hamiltonian~(\ref{eq:H_tot}) and the partition function~${\cal Z}_{\rm IM}$ of the corresponding ferromagnetic spin-$1/2$ Ising lattice given by the Hamiltonian ${\cal H}_\mathrm{IM}= -J_{eff}\sum_{\langle j,n\rangle}^{Nq/2} \mu_{\!j}^z\mu_{n}^z$:
\begin{equation}
\label{eq:ZZI}
{\cal Z} = A^{Nq/2}{\cal Z}_{\rm IM}.
\end{equation}
Note that the partition function of the uniform ferromagnetic spin-$1/2$ Ising model has exactly been calculated for various archimedean lattices~\cite{Ons44,Hou50,Dom60,McCoy73,Lav99}. From this point of view, the mapping relation~(\ref{eq:ZZI}) formally closes the rigorous treatment of the proposed 2D spin-electron model.

\subsection{Basic thermodynamic quantities}
\label{subsec:2.3}

The mapping relation~(\ref{eq:ZZI}) allows one to rigorously examine all basic thermodynamic quantities of the considered model regardless of its lattice topology (the number $q$ of the corner-sharing bipyramids). Specifically, the numerical results for the Helmholtz free energy ${\cal F}$ and, subsequently, also for the entropy ${\cal S}$ and/or the specific heat ${\cal C}$ can be obtained by means of the following fundamental thermodynamic relations:
\begin{equation}
\label{eq:FSC}
{\cal F}=-k_{\rm B}T\ln{\cal Z},
\quad
{\cal S} = -\frac{\partial {\cal F}}{\partial T}, 
\quad 
{\cal C} = -T\frac{\partial^2 {\cal F}}{\partial T^2}.
\end{equation}
Furthermore, by combining the relation~(\ref{eq:ZZI}) with exact mapping theorems developed by Barry {\it et al.}~\cite{Bar88,Kha90,Bar91,Bar95} and the generalized Callen-Suzuki identity~\cite{Cal63,Suz65,Bal02}, rigorous analytical formulas for the spontaneous sub-lattice magnetization per localized Ising spin ($m_{\rm I}$) and half-filled electron triangular cluster ($m_{\rm e}$) can also be derived:
{\setlength\arraycolsep{2pt}
\begin{eqnarray}
\label{eq:mI}
m_{\rm I} &\equiv& \big\langle\hat{\mu}_{\!j}^z\big\rangle = \big\langle\mu_{\!j}^z\big\rangle_{\!J_{eff}} = m_{\rm IM}, 
\\
\label{eq:me}
m_{\rm e} &\equiv& \big\langle \hat{S}_{\!\!j}^z\big\rangle = -\frac{2m_{\rm IM}}{w_1}\Bigg\{3\sinh\left(\frac{3\beta J}{2}\right) + 
\sinh\left(\frac{\beta J}{2}\right)\bigg[ 1+2{\rm e}^{-\beta U}
\nonumber \\
&&
\hspace{1cm}+\,2{\rm e}^{-2\beta\left(U - \sqrt{U^2+27t^2}\,\right)/3}\sum_{r=1}^{3}\cos\left(\frac{\varphi_r}{3}\right)\bigg]\Bigg\}.
\end{eqnarray}}
In above, the symbols $\langle\ldots\rangle$ and $\langle\ldots\rangle_{\!J_{eff}}$ denote the canonical averages performed over the investigated mixed spin-electron model and the effective spin-$1/2$ Ising model on the underlying $q$-coordinated archimedean lattice with the temperature-dependent but ferromagnetic exchange coupling $J_{eff}$ given by Eq.~(\ref{eq:AJef}), respectively. The magnetization  $m_{\rm IM}$ is the spontaneous single-site magnetization of the effective Ising model. We note that there are known exact solutions of $m_{\rm IM}$ for several regular spin-$1/2$ Ising lattices~\cite{Dom60,Yan52,Pot52,Nay54,Bar91}. 

In view of the above notation, the total spontaneous magnetization $m$ of the spin-electron model on a bond-decorated $q$-coordinated planar lattice normalized per bipyramidal plaquette can be defined as:
\begin{equation}
\label{eq:m}
m = \frac{2m_{\rm I} + qm_{\rm e}}{q}.
\end{equation}

\subsection{Critical temperature}
\label{subsec:2.4}

Last but not least, the rigorous criterion determining critical temperature of the considered 2D spin-electron model also follows from the mapping relation~(\ref{eq:ZZI}). Namely, the formula~(\ref{eq:ZZI}) clearly indicates that the mixed spin-electron system defined through the Hamiltonian~(\ref{eq:H_tot}) may exhibit critical behavior only if the effective ferromagnetic spin-$1/2$ Ising model to which it has been mapped is at a critical point. Thus, the critical temperature of the investigated model can be straightforwardly obtained by setting the effective coupling $J_{eff}$ listed in Eq.~(\ref{eq:AJef}) to its critical value for the corresponding archimedean lattice. In this regard, we refer the reader to Ref.~\citen{Str15}, where exact critical temperatures of a few regular and semi-regular ferromagnetic spin-$1/2$ Ising archimedean lattices are summarized in Tab.~4.1 on page~332.

\section{Discussion of numerical results}
\label{sec:3}

This section involves a discussion of the most interesting numerical results for the proposed mixed spin-electron model. Although the outcomes derived in the foregoing section are valid for both the ferromagnetic ($J>0$) and antiferromagnetic ($J<0$) Ising-type interaction, the analysis will be restricted to the ferromagnetic version of the model with $J>0$. The reason is that the sign transformation $J\to -J$ causes just a trivial sign change of the Ising spin states with respect to their electron neighbors. All ground-state and finite-temperature properties of the model remain qualitatively unchanged. 
Taking into account the aforementioned sign restriction one may set in the following the ferromagnetic Ising-type interaction $J>0$ as an energy unit to define the dimensionless quantities $t/J$,  $U/J$, and $k_\mathrm{B}T/J$ measuring relative strengths of the hopping integral, Coulomb repulsion, and temperature, respectively. 

\subsection{Ground state}
\label{subsec:3.1}

First we take a closer look at possible magnetic ground-state arrangement of the model, which can be determined by a systematic inspection of twenty eigenvalues~(\ref{eq:E_j}) of the plaquette Hamiltonian~(\ref{eq:H_j}) for all spin state combinations of the Ising spin pair $\mu_{j}^{z}$, $\mu_{j+1}^{z}$ thanks to a validity of the first commutation relation in Eq.~(\ref{eq:commutations}). In this way, two different spontaneously long-range ordered phases can be identified as energetically favorable ground states. One is the classical ferromagnetic (CFM) phase with all Ising spins in the state $\mu_j^z = 1/2$ and the highest possible total spin $S_{\!\!j}^z=3/2$ of the electron triangular clusters: 
\begin{equation}
\label{eq:CFM}
|{\rm CFM}\rangle = \prod_{j=1}^{Nq/2}
        \left|\,\mu_j^z = \frac{1}{2}\right\rangle_{\!\!j}\otimes \left|\,S_{\!\!j}^z=\frac{3}{2}\right\rangle_{\!\triangle_j}.
\end{equation}
The other one is the quantum ferromagnetic (QFM) phase, where the Ising spins again occupy the spin state $\mu_j^z = 1/2$, while the total spin of the electron triangles takes the reduced value $S_{\!\!j}^z=1/2$:
\begin{equation}
\label{eq:QFM}
|{\rm QFM}\rangle =   \prod_{j=1}^{Nq/2}
        \left|\,\mu_j^z = \frac{1}{2}\right\rangle_{\!\!j}\otimes \left|\,S_{\!\!j}^z=\frac{1}{2},\, R \textrm{ or } L\right\rangle_{\!\triangle_j}.
\end{equation}
The state vector $|S_{\!\!j}^z = 3/2\rangle_{\triangle_j}$ appearing in the eigenvector~(\ref{eq:CFM}) refers to a perfect ferromagnetic spin arrangement of the mobile electrons in the $j$-th electron triangle, 
\begin{equation}
\left|S_{\!\!j}^z=\frac{3}{2}\right\rangle_{\!\triangle_j}\!\! = |\uparrow,\uparrow,\uparrow\rangle_{j}, \nonumber
\end{equation} 
while $|S_{\!\!j}^z = 1/2, R \textrm{ or } L\rangle_{\triangle_j}$ in~Eq.~(\ref{eq:QFM}) involves the quantum superposition of nine different spin states with an opposite orientation of one mobile electron with respect to the rest two in the appropriate electron triangle including $L$eft- and $R$ight-hand side chirality of the cluster:
{\setlength\arraycolsep{2pt}
\begin{eqnarray*}
\left|\,S_{\!\!j}^z=\frac{1}{2},L\right\rangle_{\!\triangle_j} &=& \delta_1^+|\uparrow,\uparrow,\downarrow\rangle_{j}\! +\delta_1^-|\downarrow,\uparrow,\uparrow\rangle_{j}\!  +\delta_3|\uparrow,\downarrow,\uparrow\rangle_{j}
\\[-3mm]
&+&\delta_2^+|\emptyset,\uparrow,\uparrow\downarrow\rangle_{j}\! + \delta_2^-|\uparrow,\uparrow\downarrow,\emptyset\rangle_{j}\!  +\delta_3|\uparrow\downarrow,\emptyset,\uparrow\rangle_{j}  
\\
&+&\delta_4\big(|\uparrow\downarrow,\uparrow,\emptyset\rangle_{j}\! 
             +\omega|\emptyset,\uparrow\downarrow,\uparrow\rangle_{j}\! 
            +\omega^2|\uparrow,\emptyset,\uparrow\downarrow\rangle_{j}\big), 
\\    
\left|\,S_{\!\!j}^z=\frac{1}{2},R\right\rangle_{\!\triangle_j} &=& \delta_1^+|\uparrow,\uparrow,\downarrow\rangle_{j}\! +\delta_1^-|\downarrow,\uparrow,\uparrow\rangle_{j}\!  +\delta_3|\uparrow,\downarrow,\uparrow\rangle_{j}
\\[-3mm]
&+&\delta_2^+|\emptyset,\uparrow,\uparrow\downarrow\rangle_{j}\! + \delta_2^-|\uparrow,\uparrow\downarrow,\emptyset\rangle_{j}\!  +\delta_3|\uparrow\downarrow,\emptyset,\uparrow\rangle_{j}  
\\
&+&\delta_4\big(|\uparrow\downarrow,\uparrow,\emptyset\rangle_{j}\! 
             +\omega^2|\emptyset,\uparrow\downarrow,\uparrow\rangle_{j}\! 
            +\omega|\uparrow,\emptyset,\uparrow\downarrow\rangle_{j}\big).
            \nonumber 
\end{eqnarray*}}
In above, $\omega = {\rm e}^{2\pi{\rm i}/3}$ (${\rm i}^2=-1$) is the cube root of unity and the coefficients $\delta_{1(2)}^\pm, \delta_3, \delta_4$, which determine a quantum entanglement of the relevant electron states, are the functions of the dimensionless parameters $t/J$ and $U/J$:
{\setlength\arraycolsep{2pt}
\begin{eqnarray*}
\delta_1^\pm &=& \pm\frac{9}{\sqrt{6}}\frac{t/J(P\!\pm\!3t/J)}{\sqrt{P^2(P\!-\!3U/J)^2\!+\!243(U/J)^2(t/J)^2\!+\!2187(t/J)^4}},
\\
\delta_2^\pm &=& \pm\frac{9}{\sqrt{6}}\frac{t/J(P\!-\!3U/J\!\pm\!3t/J)}{\sqrt{P^2(P\!-\!3U/J)^2\!+\!243(U/J)^2(t/J)^2\!+\!2187(t/J)^4}},
\\
\delta_3 &=& -\frac{54}{\sqrt{6}}\frac{(t/J)^2}{\sqrt{P^2(P\!-\!3U/J)^2\!+\!243(U/J)^2(t/J)^2\!+\!2187(t/J)^4}},
\\
\delta_4 &=&  \frac{1}{\sqrt{6}}\frac{P(P\!-\!3U/J) -\! 27(t/J)^2}{\sqrt{P^2(P\!-\!3U/J)^2\!+\!243(U/J)^2(t/J)^2\!+\!2187(t/J)^4}}.
\end{eqnarray*}}
Here, we have introduced the new parameter $P = U/J + 2\sqrt{(U/J)^2+27(t/J)^2}\cos\phi_{U/J,t/J}$ to partially shorten relatively extensive expressions of the $\delta$ coefficients.

It should be noted that both the classical and quantum spin-electron arrangements described by the eigenvectors~(\ref{eq:CFM}) and~(\ref{eq:QFM}), respectively, have their energetically equivalent variants with the negative Ising spin $\mu_j^z = -1/2$ and the negative total spin $S_{\!\!j}^z = -3/2$ ($-1/2$) of the electron triangular clusters. These can be easily obtained by flipping all the Ising spins and electrons in individual ket vectors. For both the spin-electron configurations, the total energies of the CFM and QFM phases are (in terms of dimensionless parameters) given by:
{\setlength\arraycolsep{2pt}
\begin{eqnarray}
\label{eq:E_FM}
E_{\rm CFM} &=&  -\frac{3Nq}{2}, \\
\label{eq:E_QFM}
E_{\rm QFM} &=& -\frac{Nq}{12}\left[3 - 4\frac{U}{J} + 4\!\sqrt{\left(\frac{U}{J}\right)^2\!\!+\!27\left(\frac{t}{J}\right)^2}\cos\phi_{U/J,t/J}\right]\!.\quad\,\,\,\,
\end{eqnarray}}
The invariance of the system against a complete simultaneous spin flipping in the electron and Ising sublattices clearly declares the two-fold degeneracy of the CFM phase and that the macroscopic degeneracy $2^{Nq/2+1}$ of the QFM phase.

A direct comparison of the ground-state energies (\ref{eq:E_FM}) and (\ref{eq:E_QFM}) leads to the following analytical condition for the ground-state boundary between observed spontaneously ordered ferromagnetic phases:
\begin{equation}
 \label{eq:FM-QFM}
3 + 2\frac{U}{J} - 2\sqrt{\left(\frac{U}{J}\right)^{2}\!+27\left(\frac{t}{J}\right)^{2}}\cos\phi_{U/J,t/J} = 0.
\end{equation}
The two-fold degenerate CFM phase is limited to the ground-state parameter region $3 + 2U/J - 2\!\sqrt{(U/J)^2+27(t/J)^2}\cos\phi_{U/J,t/J} > 0$. Otherwise, the macroscopically degenerate QFM phase represents the ground state. The location of both the phases in the dimensionless zero-temperature parameter plane $t/J-U/J$ is more obvious from the shifted ground-state phase diagram shown in Fig.~\ref{fig:2}. 

\subsection{Finite-temperature phase diagrams}
\label{subsec:3.2}

\begin{figure}[t!]
\vspace{0.5cm}
\centering
  \includegraphics[width=1.0\columnwidth]{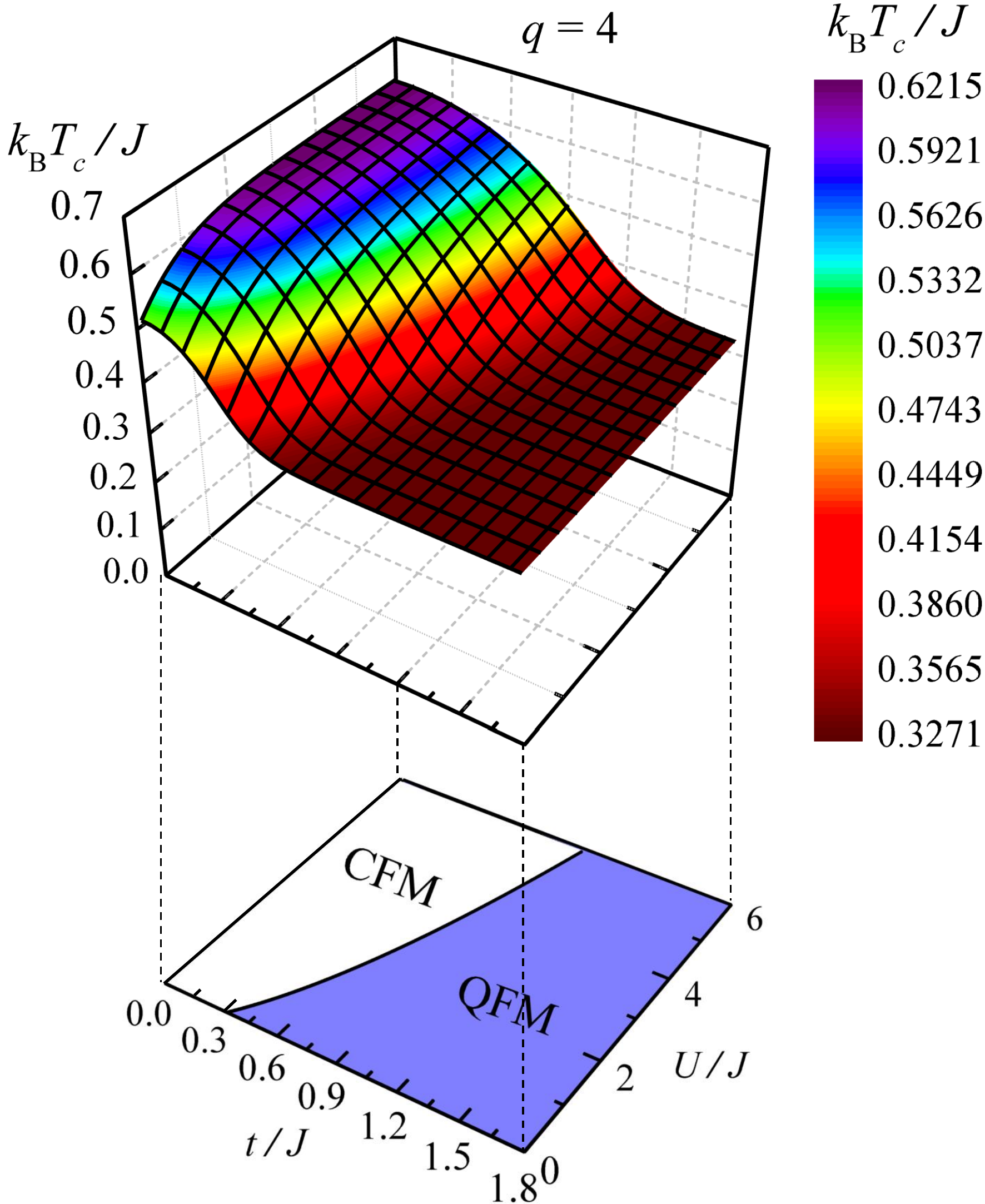}
\vspace{-4mm}
\caption{(Color online) 3D view of the finite-temperature phase diagram of the mixed spin-electron model on a bond-decorated square lattice which is schematically illustrated in Fig.~\ref{fig:1}. The plotted phase diagram is supplemented with shifted zero-temperature $t/J-U/J$ parameter plane showing the ground-state phase diagram of the model.}
\label{fig:2}
\end{figure}

\begin{figure}[t!]
\vspace{0.5cm}
\centering
\includegraphics[width=1.0\columnwidth]{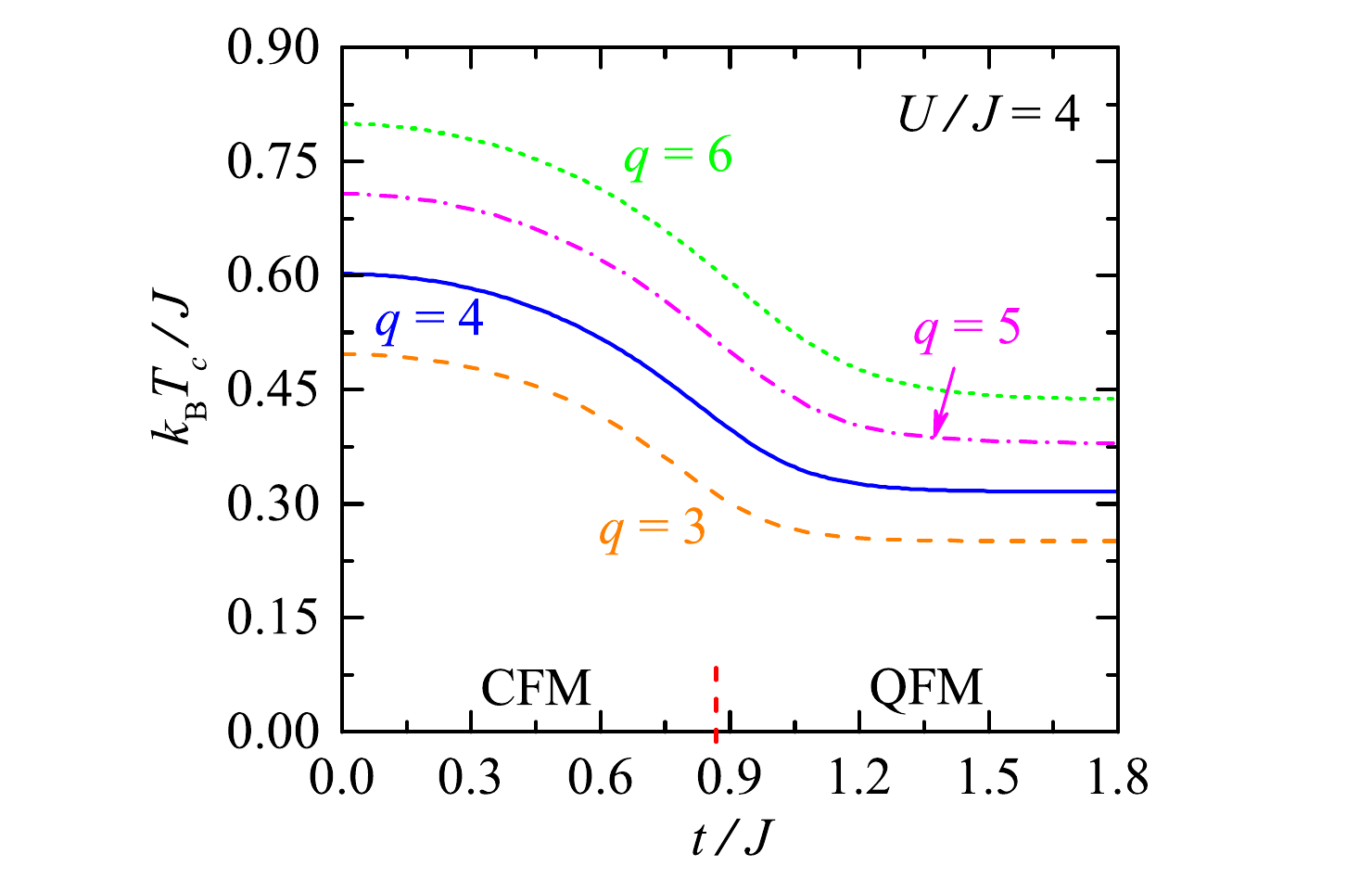}
\vspace{-0.5cm}
\caption{(Color online) Critical temperatures of the mixed spin-electron model on bond-decorated honeycomb ($q=3$), kagom\'e  ($q=4$), trellis ($q=5$), and triangular ($q=6$) lattices as a function of the hopping parameter $t/J$ at the fixed Coulomb term $U/J=4$. The vertical black dashed line crossing the horizotal axis labels the position of the ground-state phase transition between the CFM and QFM phases.}
\label{fig:3}
\end{figure}

In order to gain insight into a temperature stability of the spontaneous  spin-electron order observed in the CFM and QFM ground states, in this subsection we will turn our attention to the finite-temperature phase diagrams that are displayed in Figs.~\ref{fig:2} and~\ref{fig:3}.  

Figure~\ref{fig:2} shows the global 3D finite-temperature phase diagram of the particular bond-decorated square lattice, where four spin-electron bipyramidal plaquettes share a common vertex occupied by the localized Ising spin (see Fig.~\ref{fig:1}). The colored surface representing the critical temperature $k_{\rm B}T_{c}/J$ of the system was numerically calculated from the well-known exact critical condition $\beta_c J_{eff} = 2\ln(1+\sqrt{2})$ ($\beta_c=1/k_{\rm B}T_c$) for the pure ferromagnetic spin-$1/2$ Ising square lattice~\cite{Ons44}. For clarity, the plotted finite-temperature diagram is supplemented with the shifted zero-temperature $t/J-U/J$ parameter plane, which represents the ground-state phase diagram of the investigated model. It can be easily understood from the figure that the spontaneous quantum ferromagnetic spin-electron arrangement with interesting local chiral degrees of freedom of the mobile electrons from the same trigonal bipyramid, which is inherent to the QFM phase, is thermally less stable than the classical one present in the CFM phase. In fact, the critical temperature of the former phase takes the constant value $k_{\rm B}T_{c}/J \approx 0.3271$ almost above its entire stability region, while that corresponding to the latter is always higher. The highest critical temperature of the CFM phase can be found for negligible values hopping terms compared to the Ising-type interaction $t/J\to 0$ in the asymptotic limit of infinitely strong Coulomb repulsion $U/J\to\infty$ (practically already at $U/J \approx 7$). In this particular region, the critical temperature of the system achieves almost double value $k_{\rm B}T_{c}/J \approx 0.6215$ than in the region corresponding to the QFM phase.

Qualitatively similar critical behavior can also be observed for many other regular and/or semi-regular bond-decorated planar lattices formed by corner-sharing spin-electron trigonal bipyramids. For illustration, Fig.~\ref{fig:3} summarizes typical dependences of the critical temperature on the hopping parameter $t/J$ for four such lattices at the fixed Coulomb term $U/J=4$, namely, the decorated honeycomb ($q=3$), kagom\'e ($q=4$), trellis ($q=5$), and triangular ($q=6$) lattices. The plotted curves have been obtained in the same way as the surface in Fig.~\ref{fig:2}, i.e., by setting the effective temperature-dependent coupling $J_{eff}$ listed in Eq.~(\ref{eq:AJef}) to the critical point of the underlying ferromagnetic spin-$1/2$ Ising honeycomb, kagom\'e, trellis, and triangular lattices $2\ln(2+\sqrt{3})$, $\ln(3+2\sqrt{3})$, $\ln 4$, and $\ln 3$, respectively~\cite{Str15}.

\subsection{Spontaneous sub-lattice and total magnetization}
\label{subsec:3.3}

\begin{figure}[t!]
\vspace{0.5cm}
\centering
\includegraphics[width=1.0\columnwidth]{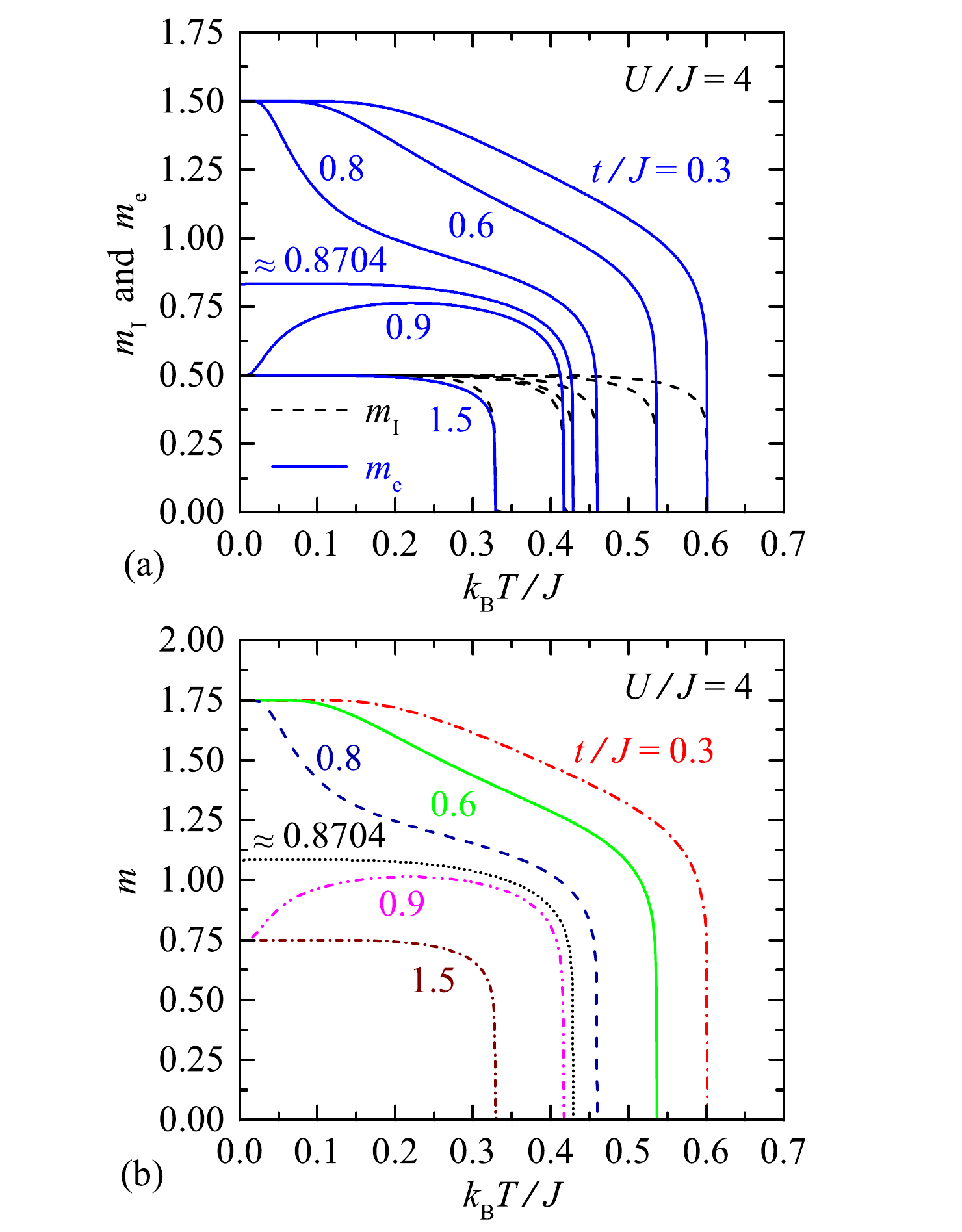}
\vspace{-4mm}
\caption{(Color online) Temperature dependences of the spontaneous sublattice magnetization $m_I$, $m_e$ [panel~(a)] and the total spontaneous magnetization $m$ per plaquette cell [panel~(b)] of the mixed spin-electron model on a bond-decorated square lattice for the fixed Coulomb term $U/J = 4$ and a few representative values of the hopping parameter $t/J$.}
\label{fig:4}
\end{figure}

The observed critical behavior of the mixed spin-electron system and its ground-state arrangement can be independently confirmed by temperature dependences of the spontaneous sublattice and total magnetization, which are plotted in Fig.~\ref{fig:4} for the fixed Coulomb term $U/J = 4$ and several representative values of the hopping parameter $t/J$. Without losing the generality of the discussion, the plotted magnetization curves correspond to the particular bond-decorated square lattice schematically shown in Fig.~\ref{fig:1}. 

In accordance with the ground-state analysis, the sublattice magnetization depicted in Fig.~\ref{fig:1}(a) for the hopping amplitudes $t/J$ smaller (higher) than the boundary value $t_b/J\approx 0.8704$ start from their asymptotic zero-temperature values $m_{\rm I}=0.5$ and $m_{\rm e}=1.5$ ($0.5$) and rapidly fall down to zero value at a critical temperature according to the power law with the critical exponent $\beta_m=1/8$ from the standard Ising universality class. However, there is an obvious difference in temperature variations of $m_{\rm I}$ and $m_{\rm e}$ at low and moderate temperatures. More specifically, the sublattice magnetization $m_{\rm I}$ remains almost constant over a relatively wide temperature range for any value $t/J$. On the contrary, the sublattice magnetization $m_{\rm e}$ falls down the more steeply with increasing temperature, or shows the more pronounced temperature-induced initial increase, the closer the hopping term $t/J$ is to the boundary value $t_{b}/J\approx 0.8704$. The observed pronounced thermal trend of the latter sublattice magnetization clearly indicates a greater temperature sensitivity of the electron sublattice compared to the Ising one, especially in a close vicinity of the ground-state phase transition between the long-range ordered CFM and QFM phases. The sensitivity of the electron sublattice to the thermal fluctuations is canceled only at the boundary value of the hopping parameter $t_b/J$, where the perfect ferromagnetic electron arrangement of the CFM phase coexists with the quantum one present in the QFM phase, and for strong enough hopping terms $t/J\gg t_b/J$ [see the curves of $m_{\rm e}$ plotted for $t_b/J\approx 0.8704$ and $t/J = 1.5$ in Fig.~\ref{fig:4}(a)]. 

As expected, the aforedescribed temperature variations of the sublattice magnetization $m_\mathrm{I}$ and $m_\mathrm{e}$ have a significant impact on temperature variations of the total spontaneous magnetization $m$ normalized per spin-electron bipyramidal plaquette, which can be considered as the relevant order parameter for the spontaneously long-range ordered CFM and QFM phases regardless of the number $q$ of corner-sharing plaquettes as well as current lattice topology. This statement is supported by Fig.~\ref{fig:4}(b), which presents temperature dependences of the total magnetization $m$ calculated from the curves of $m_\mathrm{I}$, $m_\mathrm{e}$ in Fig.~\ref{fig:4}(a) by using Eq.~(\ref{eq:m}). Obviously, all plotted dependences of $m$ faithfully follow temperature variations of the spontaneous sublattice magnetization $m_{\rm e}$ including the sharp initial temperature-induced decline and rise observable for $t/J\lesssim t_b/J$ and $t/J\gtrsim t_b/J$, respectively. The full saturation value $m = 1.5$ detected in the zero-temperature asymptotic limit for the hopping parameters $t/J<t_b/J$ reflects a perfect ferromagnetic spin alignment of both the Ising and electron sublattices, while the reduced one $m = 0.75$ observable for $t/J>t_b/J$ agrees with the quantum superposition of nine different configurations of the electron triangular clusters with the reduced total spin $S_{\!\!j}^z=1/2$. The nontrivial zero-temperature asymptotic value $m \approx 1.0807$ corresponding to $t_{b}/J\approx 0.8704$ clearly points to the mixture of the electron arrangements present in the CFM and QFM ground states along the phase boundary between these phases.

\subsection{Entropy and specific heat}
\label{subsec:3.4}

Our investigation of the magnetic properties of the ferromagnetic mixed spin-electron model on planar lattices formed by corner-sharing bipyramids will be completed by examining temperature variations of the entropy and specific heat. Without losing its generality, the discussion will again be limited to the particular model on the decorated square lattice, because it qualitatively captures all important features that would be observed also in another regular as well as semiregular archimedean lattices with the same bond decoration.  

\begin{figure}[t!]
\vspace{0.5cm}
\centering
\includegraphics[width=1.0\columnwidth]{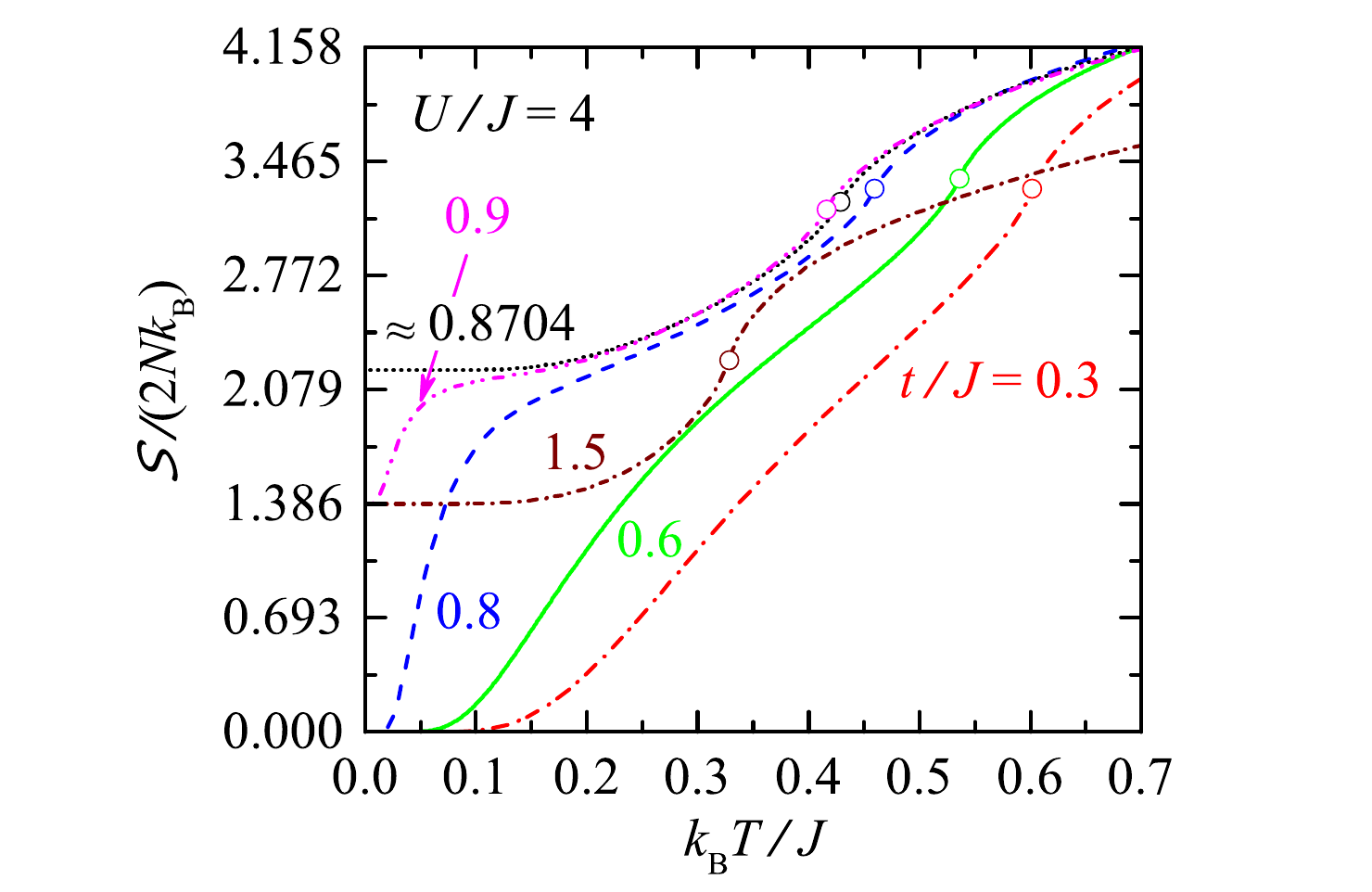}
\vspace{-4mm}
\caption{(Color online) Temperature variations of the entropy  ${\cal S}/(2Nk_{\rm B})$ per bipyramidal plaquette of the mixed spin-electron model on the same lattice and for the same values of the interaction parameters $U/J$, $t/J$ as used in Fig.~\ref{fig:4}. Empty circles label weak energy-type singularities of the physical quantity appearing at appropriate critical temperatures.}
\label{fig:5}
\end{figure}

Figure~\ref{fig:5} illustrates typical temperature dependences of the entropy ${\cal S}/(2Nk_{\rm B})$ normalized per bipyramidal plaquette. In this figure, the values of the interaction parameters $U/J$ and $t/J$ were selected so that the plotted curves are directly comparable to the temperature variations of the spontaneous magnetization shown in Fig.~\ref{fig:4}.
In general, the entropy ${\cal S}/(2Nk_{\rm B})$ rises monotonously from three different zero-temperature asymptotic values with increasing of temperature and shows a standard weak energy-type singularity (empty circle) at the continuous order-disorder phase transition. The lowest asymptotic value ${\cal S}/(2Nk_{\rm B}) = 0$, which can be observed for the hopping terms $t/J<t_b/J$, confirms the stability of the two-fold degenerate CFM phase. On the other hand, the residual one ${\cal S}/(2Nk_{\rm B}) = 2\ln 2\approx 1.386$, detected in the zero-temperature limit under the reverse condition $t/J>t_{b}/J$, is in accordance with the macroscopic degeneracy of the QFM phase caused by two possible chiral degrees of each electron triangle of the bipyramidal plaquette. Finally, the highest entropy ${\cal S}/(2Nk_{\rm B}) = 2\ln 3\approx 2.197$, which can be found at the boundary hopping term $t_b/J\approx0.8704$, reflects the macroscopic degeneracy of the model along the ground-state boundary between the spontaneously ordered CFM and QFM phases. As expected, the most rapid low-temperature rise of the entropy can be detected when the hopping term $t/J$ is selected close enough to the ground-state phase transition CFM--QFM (see the curves plotted in Fig.~\ref{fig:5} for $t/J = 0.8$ and $0.9$). In this particular region, the zero-temperature energies of the CFM and QFM phases take very close values, and therefore, both the spin-electron arrangements become accessible already at a slight temperature increase. The thermally induced mixing of the classical and quantum ferromagnetic spin-electron arrangements near CFM--QFM has already been demonstrated in more detail in Subsec.~\ref{subsec:3.3} by temperature variations of the spontaneous sublattice magnetization $m_e$ and the total magnetization~$m$ (see again the magnetization curves plotted for $t/J = 0.8$, $0.9$ in Fig.~\ref{fig:3} and compare them with the corresponding entropy ones in Fig.~\ref{fig:5}).    
\begin{figure}[t!]
\vspace{0.5cm}
\centering
\includegraphics[width=1.0\columnwidth]{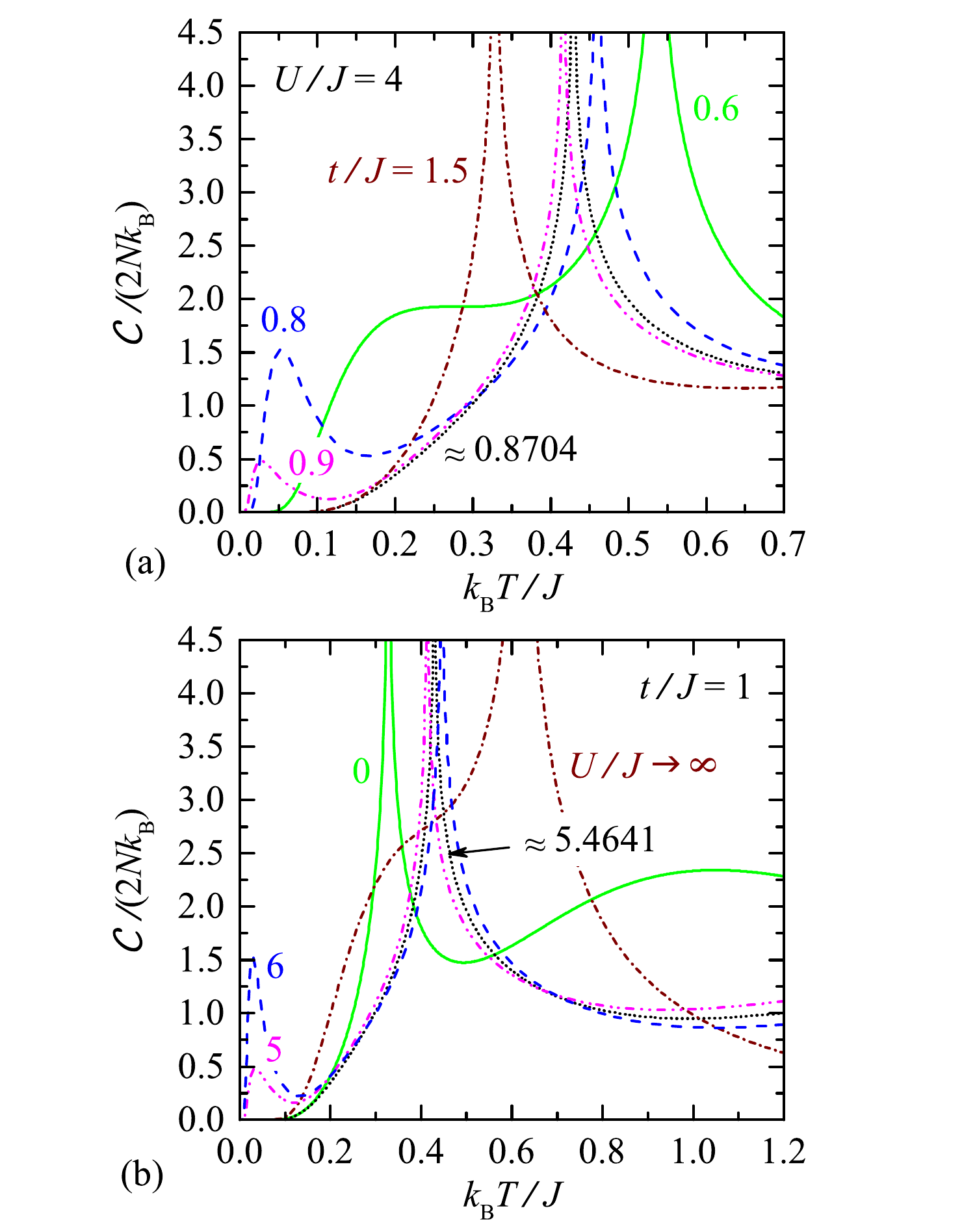}
\vspace{-4mm}
\caption{(Color online) Temperature variations of the specific heat  ${\cal C}/(2Nk_{\rm B})$ per bipyramidal plaquette of the mixed spin-electron model on a bond-decorated square lattice for the 
the fixed Coulomb term $U/J=4$ and the same values of the hopping term $t/J$ as used in Figs.~\ref{fig:4} and~\ref{fig:5} [panel (a)], and for the fixed hopping parameter $t/J = 1$ and a few representative values of the Coulomb term~$U/J$ [panel (b)].}
\label{fig:6}
\end{figure}

Last but not least, let us explore the typical temperature variations of the specific heat ${\cal C}/(2Nk_{\rm B})$ per plaquette which are depicted in Fig.~\ref{fig:6}. Figure~\ref{fig:6}(a) demonstrates the effect of the hopping parameter $t/J$, while  Fig.~\ref{fig:6}(b) shows what happens when the on-site Coulomb repulsion $U/J$ increases from zero to infinity. Obviously, all plotted specific heat curves have a single logarithmic divergence from the standard Ising universality class, which reflects the continuous order-disorder phase transition. In accordance to the critical behavior of the system, position of this divergence is gradually shifted to lower temperatures with the increasing hopping term $t/J$ [see Fig.~\ref{fig:6}(a)], and, oppositely, to the higher temperatures with increasing $U/J$ [see Fig.~\ref{fig:6}(b)]. Moreover, if the Coulomb repulsion $U/J$ and the hopping parameter $t/J$ are taken from the proximity of the ground-state phase transition CFM--QFM, an interesting low-temperature Schottky-type maximum appears in the specific heat curves in addition to the standard high-temperature one, which is more or less visible depending on the values of $U/J$ and $t/J$. A comparison of the specific heat curves plotted in Fig.~\ref{fig:6}(a) for $t/J = 0.8$, $0.9$ with corresponding 
temperature dependences of the spontaneous magnetization and entropy shown in Figs.~\ref{fig:4} and~\ref{fig:5}, respectively, clearly indicates that these peaks originate from strong thermal excitations of the electron sublattice from its ground-state configuration to the low-lying excited one with the character of the neighboring phase. More specifically, the higher low-temperature maxima reflect excitations of the mobile electrons from the classical two-fold degenerate ferromagnetic arrangement to the macroscopically degenerate quantum one with local chiral degrees of freedom of these particles, while those with approximately three times lower amplitudes are results of the opposite scenario.

\section{Conclusions}
\label{sec:4}

The present work offers an insight into magnetic properties of a mixed spin-electron model on decorated planar lattices composed of half-filled trigonal bipyramids, which is exactly solved by the generalized decoration-iteration transformation establishing a rigorous mapping correspondence with the effective spin-$1/2$ Ising model on a corresponding regular or semiregular archimedean lattice. In particular, the possible ground-state configuration and finite-temperature phase diagrams have been comprehensively studied for various lattice topologies by assuming the ferromagnetic exchange coupling between the localized Ising spins and their nearest electron neighbors. The obtained numerical results have been complemented by a detailed analysis of the temperature variations of the spontaneous magnetization, entropy and specific heat. 

It has been demonstrated that the investigated spin-electron
model exhibits two spontaneously ordered ferromagnetic ground states irrespective of its lattice topology. One is the two-fold degenerate classical phase with the perfect ferromagnetic spin order inside and also between Ising and electron sublattices, while the other one is the macroscopically degenerate quantum phase characterized by two possible chiral degrees of freedom of the mobile electrons hopping within the same bipyramidal plaquette. It has been shown that the spontaneous classical spin-electron arrangement is for some combinations of model's parameters thermally almost twice more stable than the quantum one. The existence of both the spin-electron arrangements and their resistance to thermal fluctuations have also been independently verified by the respective temperature dependences of the spontaneous sublattice and total magnetization, as well as entropy and specific heat.    

\begin{acknowledgment}
\section*{Acknowledgments}
This work was financially supported by the grant of Slovak Research and Development Agency under the contract No. APVV-20-0150 and by The Ministry of Education, Science, Research and Sport of the Slovak Republic under the grands Nos. VEGA 1/0105/20 and KEGA 021TUKE-4/2020.
\end{acknowledgment}

\end{document}